\documentclass[a4paper,english]{article}
\usepackage[T1]{fontenc}
\usepackage[latin9]{inputenc}
\usepackage{float}
\usepackage{units}
\usepackage{graphicx}
\usepackage{esint}
\usepackage{babel}
\usepackage{xcolor}
\usepackage{subfig}
\usepackage{amssymb}
\usepackage{placeins}

\makeatletter

\pdfpageheight\paperheight
\pdfpagewidth\paperwidth

\usepackage{caption}
\usepackage{prettyref}
\newrefformat{sfig}{Supplementary Figure \ref{#1}}
\DeclareCaptionLabelFormat{support}{Supplementary #1 #2} 

\makeatother

\newcommand{\nexp}{N_{\mathrm{exp}}}
\newcommand{\thalf}{t_{\nicefrac{1}{2}}}
\newcommand{\tc}{t_{\mathrm{c}}}

\newcommand{\pexp}{P_{\mathrm{exp}}}
\newcommand{\fsim}{f_{\mathrm{th}}}
\newcommand{\fexp}{f_{\mathrm{exp}}}
\newcommand{\ri}{R}
\newcommand{\rt}{r}
\newcommand{\friction}{\xi}
\newcommand{\gnre}{\gamma}
\newcommand{\gre}{\tilde \gamma}
\newcommand{\dg}{D_{\gamma}}
\newcommand{\snre}{\sigma}
\newcommand{\sre}{\tilde \sigma}
\newcommand{\ds}{D}
\newcommand{\DP}[2]{\ensuremath{\frac{\partial{#1}}{\partial{#2}}}}
\newcommand{\DPn}[3]{\ensuremath{\frac{\partial^{#1}{#2}}{\partial{#3^{#1}}}}}
\newcommand{\DT}[2]{\ensuremath{\frac{\mathrm{d}{#1}}{\mathrm{d}{#2}}}}
\newcommand{\DTn}[3]{\ensuremath{\frac{\mathrm{d}^{#1} {#2}}{\mathrm{d} {#3}^{#1} }}}

\begin{document}

\begin{center}
  
{\Large \bf Tissue fusion over non-adhering surfaces}

\bigskip
Vincent Nier, Maxime  Deforet, Guillaume  Duclos, Hannah   G. Yevick,
Olivier   Cochet-Escartin, Philippe Marcq$^*$ and Pascal Silberzan$^{\dagger}$

\bigskip
Laboratoire Physico-Chimie Curie,
Institut Curie, Universit\'e Pierre et Marie Curie,\\ 
CNRS UMR 168, 75005 Paris, France

$^*$ philippe.marcq@curie.fr 
$^{\dagger}$pascal.silberzan@curie.fr
\end{center}

\section*{Abstract}

Tissue fusion eliminates physical voids in a tissue  to  form  a  continuous
structure and is central  to  many  processes  in  development  and  repair.
Fusion events in vivo, particularly in embryonic development, often  involve
the purse-string contraction of a  pluricellular  actomyosin  cable  at  the
free edge. However in vitro,  adhesion  of  the  cells  to  their  substrate
favors a closure mechanism mediated by lamellipodial protrusions, which  has
prevented a systematic study of the purse-string mechanism.  Here,  we  show
that monolayers can  cover  well-controlled  mesoscopic  non-adherent  areas
much larger than a  cell  size  by  purse-string  closure  and  that  active
epithelial fluctuations are required for this process. We have formulated  a
simple stochastic model that  includes  purse-string  contractility,  tissue
fluctuations and effective  friction  to  qualitatively  and  quantitatively
account for the dynamics of closure. Our data suggest that, in vivo,  tissue
fusion  adapts  to  the  local  environment  by  coordinating  lamellipodial
protrusions and purse-string contractions.

\section*{Introduction}
\label{sec:itd}

Tissue fusion is a frequent and important  event  during  which  two  facing
identical tissues meet and bridge collectively over  a  gap  before  merging
into a continuous  structure  (1).  Imperfect  tissue  fusion  in  embryonic
development results in congenital defects for instance, in the  palate,  the
neural  tube  or  the  heart  (1).  Epithelial  wound  healing  is   another
illustration of tissue fusion through which a gap in  an  epithelium  closes
to restore the integrity of the monolayer (2).

Model in vitro experiments have been  developed  using  cell  monolayers  to
study the different stages of healing from collective cell migration to  the
final stages of closure. In this context, we (3)  and  others  (4,  5)  have
recently demonstrated that, for  cells  adhering  to  their  substrate,  and
despite the presence of a contractile peripheral  actomyosin  cable  at  the
free edge, the final stages of closure of wounds larger than a typical  cell
size result mostly from protrusive lamellipodial activity at the border.  In
that case, the function of the  actin  cable  appears  to  be  primarily  to
prevent the onset of migration fingers led by leader cells (6) at  the  free
edge. Cell crawling has also been shown to  have  a  major  role  in  tissue
fusion in vivo, for example during the closure of epithelial wounds  in  the
Drosophila embryo (7).

However, in  physiological  developmental  situations,  there  is  often  no
underlying substrate to which lamellipodia  can  adhere  to  exert  traction
forces. This is the case, for instance, in neural tube formation (8)  or  in
wound healing  in  the  Xenopus  oocyte  (9).  The  generally  well-accepted
mechanism in these adhesion-free situations is  the  so-called  purse-string
mechanism in which the actomyosin cable at the edge of the  aperture  closes
it by  contractile  activity  (10).  Note  that  the  purse-string  and  the
crawling mechanisms are not mutually exclusive (11) and may be  involved  at
different stages of the closure (5, 12). In  addition,  "suspended"  cohorts
of cells, which do not interact with a substrate besides being  anchored  to
a few discrete attachment points, are also observed in  situations  such  as
collective migration in cancer invasion (13).

Several experimental studies have  documented  protrusion-driven  collective
migration in vitro, but the purse-string mechanism has not  been  thoroughly
investigated in model situations. Such an analysis imposes to  suppress  the
contribution of the protrusions to closure and, therefore,  to  conduct  the
experiments on non-adherent substrates.

In a seminal paper, fibroblast sheets were shown to grow  and  migrate  with
their sides  anchored  to  thin  glass  fibers  (14).  More  recent  studies
extended this observation  to  keratinocyte  monolayers  or  epidermal  stem
cells  bridging  between  microcontact-printed  adhesive  tracks  (15,  16).
However, despite recent advances emphasizing the role of  tissue  remodeling
(17), the mechanism of closure of a suspended epithelium in the  absence  of
these anchoring sites remains an open question. To address  this  point,  we
have studied the dynamics of gap closure in  an  unsupported  epithelium  in
which the actomyosin cable and the suspended tissue could not adhere to  the
substrate. Purse-string contractility in  the  absence  of  protrusions  was
therefore studied on well-defined mesoscopic non-adherent patches within  an
adherent substrate.

\begin{figure}[!t]
\centering
\includegraphics[scale=0.6]{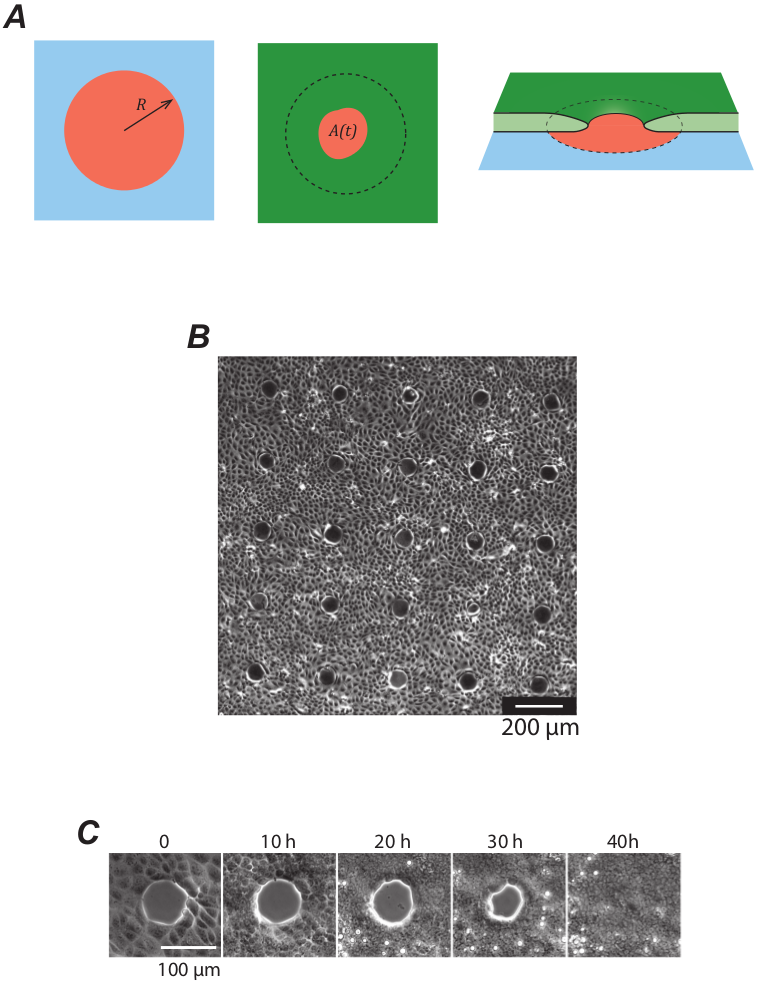}
\caption{
\textbf{Fusion of an epithelium over non-adhering domains.}
\textbf{A} Schematics of the experiment: the surface (blue) is patterned 
with non-adhering
domains (red) of radius $R$. Cells (green) progressively cover these domains
and the cell-free area ($A(t)$) is dynamically monitored.
\textbf{B} $5 \times 5$ array of non-
adhering domains in phase contrast at the onset of the experiment 
($R = 42 \, \mu \mathrm{m}$).
\textbf{C} Time evolution of a single domain ($R = 42 \, \mu \mathrm{m}$).
}
\end{figure}

\section*{Results}
\label{sec:res}

We studied the bridging of a monolayer over a well-defined non-adherent  gap
on adherent glass substrates patterned with strictly  non-adhesive  circular
regions of radius $R$ between $5$ and $75\,\mu$m (Figure 1A). The  surface  
treatment
kept its non-adherent properties for up to $3$  weeks  in  biological  buffers
(18, 19). To ensure we obtained reliable statistics, we worked  with  arrays
of tens of non-adherent identical domains on which  we  cultured  epithelial
Madin Darby Canine Kidney (MDCK) cells (Figure 1B) (20). Notably, MDCK  cell
sheets have been previously shown to remain functional when  suspended  over
large distances in culture medium (21).  Fusion  processes  in  neighbouring
domains remained independent by imposing a space between each  of  at  least
$300\,\mu$m, a distance larger than  the  velocity  correlation  length  
measured
independently  for  the  same  cellular  system  (22).  Tissue  fusion   was
monitored from confluence ($t = 0$) up to $4$ days.

Immediately after seeding, the cells  adhered  on  the  glass  and  colonies
developed by proliferation. The expanding  monolayer  readily  covered  non-
adhesive domains that had a radius of less than $10\,\mu$m. In these  cases,  the
advancing front edge made no arrest, confirming that cells have the  ability
to bridge over non-adherent defects smaller  than  their  own  size  (23-26)
(Supp. Figure 1). At the other extreme, for domains with  a  radius  greater
than $70\,\mu$m, the monolayer covered only the  glass  surface  surrounding  the
patches (Supp. Figure 1). After several days, we  observed  the  development
of a tridimensional "rim" at  the  boundaries  of  the  domains  as  already
reported (19) but no further evolution in the subsequent weeks (19, 27).

Between these two limiting situations, the  monolayer  initially  surrounded
the  non-adhesive  domains  and  then  proceeded  to  cover  them  until  it
eventually fused (Figure 1C, Supp. Figure 1, Supp.  Video  1).  Observations
using confocal microscopy at the non-adherent surface / monolayer  interface
revealed the absence of vinculin or paxillin, two proteins  associated  with
focal adhesions. This confirmed our basic  assumption:  the  cells  did  not
develop adhesions with the treated surface during and after  closure  (Supp.
Figure 2).

We followed the closure process by monitoring the area covered by  cells  on
domains of various sizes over  a  period  of  several  days.  A  significant
fraction of the domains with radii less than $30\,\mu$m were already closed  when
the monolayer reached confluence. Therefore, the mechanism  by  which  cells
cover these small domains may be different (for example by direct  bridging)
from the one relevant to larger domains. As a consequence,  we  limited  our
study to $30 \, \mu\mathrm{m} < R < 50\,\mu\mathrm{m}$.  The  cell-free  area  
$A(t)$  showed  only  minor
distortions to a quasi-circular shape, allowing us to  define  an  effective
radius $r(t)$ as $r(t) = \sqrt{A(t)/\pi}$ (Figure 1A).

It  is  worth  comparing  the  present  experiments  with  the  healing   of
comparable size wounds  of  the  same  cell  line  on  homogeneous  adhesive
substrates in which protrusions at the leading edge were  shown  to  be  the
driving force for closure (3, 4). In both  cases,  the  shape  of  the  hole
remained relatively circular; in particular, no  fingering  of  the  leading
edge (6) was observed. However,  the  absence  of  cell-substrate  adhesions
drastically slowed down the closing dynamics (typically $30$ h in the  present
setting vs. $3$ h on a homogeneously  adherent  substrate  for  $R = 35\,\mu$m).
Moreover, in the experiments described here, the closing  was  very ``noisy''
in two  respects.  First,  a  given  hole  closed  in  a  seemingly  erratic
succession of large amplitude retractions and expansions of  the  open  area
(Figure 2A). In some experiments, we observed closure down  to  $20\,\%$  of  the
initial radius, which then re-opened to  $50\,\%$  before  eventually  closing.
However, once the closure was fully completed, there was no re-opening  (and
no indication of a different morphology of the cells over  the  non-adhesive
patch compared to the  adhesive  surface  (Figure  1C)).  Second,  comparing
several closure events for the same non-adhesive patch size, we  observed  a
very large dispersion of the closure times (Supplementary Figure  3,  Figure
2C-G). For instance, if $R = 35\,\mu$m, the average closing time  was  $44$  h  and
the standard deviation was $18$ h ($N=150$). Because of this  large  dispersion,
the entire distributions of the  closure  times  (and  therefore  meaningful
average closure times) after $83$ h could be accessed only for patches with  a
radius less than  $35 \,\mu$m.  Unfortunately,  the  development  of  the  above-
mentioned 3D rim at the border after typically  $4$  days  prevented  us  from
accessing the long-time parts  of  these  distributions  for  larger  domain
sizes.

Given this large variability, we chose to reason in terms  of  the  fraction
$f(R,t)$ of closed holes at a time  $t$  for  a  given  initial  radius  $R$.  This
fraction $f$ is plotted as a function of  $R$  after  $4$  days  (Figure  2H).  As
previously mentioned, all patches with a  radius  less  than  $35\,\mu$m  closed
within $83$  h.  By  contrast,  only  a  small  fraction  of  the  experiments
performed at $R > 55\,\mu$m closed in this time-frame. As a matter  of  fact,  we
never observed the closing of patches whose radius was larger  than  $70\,\mu$m.
The full dynamic evolution of these fractions is plotted as a  heat  map  in
Figure 3A for $30 \,\mu\mathrm{m} < R < 50 \,\mu\mathrm{m}$ and 
$0 < t < 83$ h.

Closure is necessarily a collective effect as cells must form  a  continuous
structure that bridges over the non-adherent surface. Indeed, by  conducting
the experiments in low calcium  conditions  that  disrupt  cadherin-mediated
cell-cell  adhesions  (19),  the  efficiency  of  closure  was  considerably
reduced (Supp. Figure 4).

 \begin{figure}[!t]
 \centering
 \includegraphics[scale=0.55]{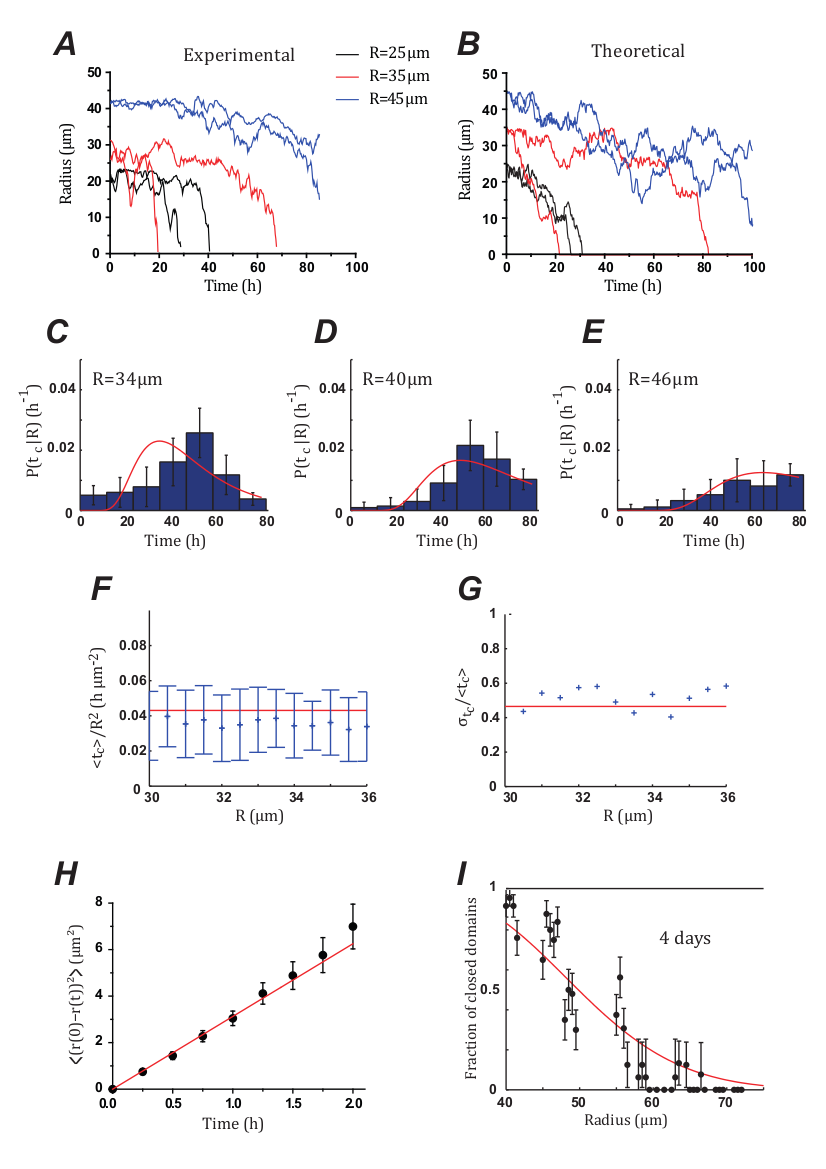}
 \caption{
 \textbf{Gap closure is a stochastic process.}
 \textbf{A-B} Comparison between
 experimental (A) and theoretical (B) trajectories. The two are visually
 very close. Note the very large fluctuations of the radius during closure.
 \textbf{C-E} The closure times are widely distributed for a given initial radius
 (bars: experimental values, lines theoretical predictions with 
 $\gre = 10\;\mu\mathrm{m^{2} h^{-1}}$ and
 $D = 1.6\;\mu\mathrm{m^{2} h^{-1}}$).
 Error bars are SEMs. $N$ values are $571$ (C), $498$ (D), $498$ (E). (F) As
 predicted by the model, the ratio $\langle t_c \rangle/R^2$ is 
 approximately constant within the accessible dynamical range
 $30 \le R \le 36 \, \mu$m, where long time closure events
 ($t_c > 84$h) are negligible. The red line corresponds to the theoretical
 prediction (Supp. Eq.~(40)) 
 $\frac{\langle \tc(\ri) \rangle}{\ri^2} = \frac{1}{2(\gre+D)}
 = 0.043 \mathrm{h} \mu\mathrm{m}^{-2} $ 
 with $\gre = 10\;\mu\mathrm{m^{2} h^{-1}}$ and
 $D = 1.6\;\mu\mathrm{m^{2} h^{-1}}$. Error bars are
 standard deviations.
 \textbf{G} The coefficient of variation of the closure time
 is approximately constant within the same range, with values consistent
 with the theoretical prediction (Supp. Eq.~(46)) indicated with a solid
 line: $ \frac{\sigma_{\tc}({\ri})}{\langle \tc(\ri) \rangle} =
  \left(\frac{2 D}{3D + \gre}\right)^{\frac{1}{2}} = 0.47$
 with $\gre = 10\;\mu\mathrm{m^{2} h^{-1}}$ and
 $D = 1.6\;\mu\mathrm{m^{2} h^{-1}}$.
 \textbf{H} Early times are well described by a diffusive process.
 Points are the experimental points resulting from the average of more than
 $400$ trajectories, the red line is a linear fit yielding a diffusion
 coefficient of $D = 1.56 \pm 0.03 \;\mu\mathrm{m^{2} h^{-1}}$  
 (error bars are SEMs).
 \textbf{I} Fractions of closed wounds at 4 days
 (black points). The red line is the theoretical fraction 
 ($\gre = 10\;\mu\mathrm{m^{2} h^{-1}}$ and
 $D = 1.6\;\mu\mathrm{m^{2} h^{-1}}$). Error  bars are SEMs.
 }
 \end{figure}

\FloatBarrier

\begin{figure}[!h]
\centering
\includegraphics[scale=0.6]{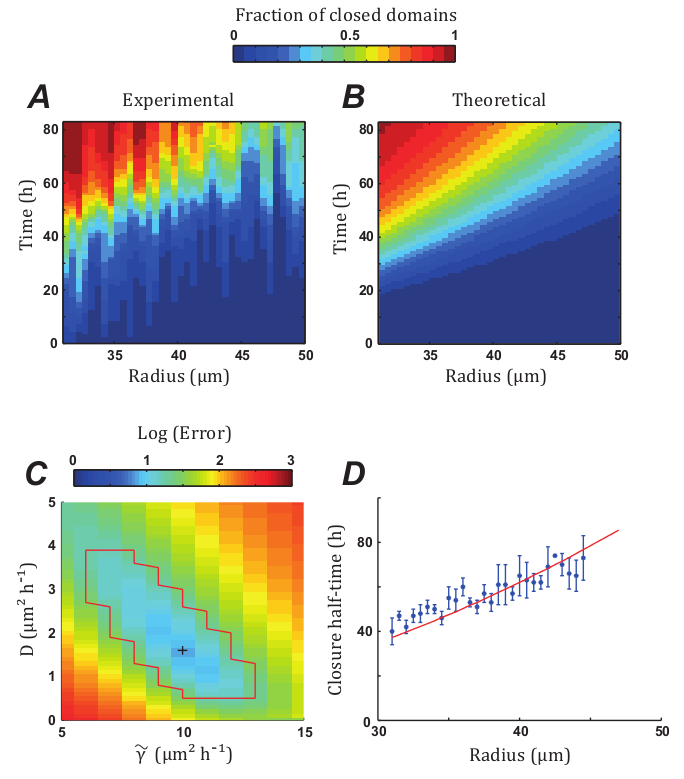}
\caption{
\textbf{The fraction of closed wounds is well described by a stochastic
model.}
\textbf{A} Experimental and
\textbf{B} theoretical fractions of the closed wounds as
a function of the initial radius $R$ and time $t$. The theoretical heatmap 
B is plotted with the values $\gre = 10\;\mu\mathrm{m^{2} h^{-1}}$ and
 $D = 1.6\;\mu\mathrm{m^{2} h^{-1}}$ 
that correspond to the smallest error in the fit of the data
(C). The two heatmaps are visually very close. \textbf{C} Plane of
the error landscape (logarithmic scale). The cross denotes the optimum, the
red contour bounds the $95 \, \%$ confidence region.
\textbf{D} The closure half-time is
the time at which half of the wounds have closed. The points are the
experimental data ($3825$ domains) and the line is the theoretical time for
$\gre = 10\;\mu\mathrm{m^{2} h^{-1}}$ and  $D = 1.6\;\mu\mathrm{m^{2} h^{-1}}$. 
The fraction of closed patches at the end of the experiment is less
than $0.5$ above $R = 45 \, \mu$m. Error bars are SEMs.}
\end{figure}

No stable lamellipodial protrusions similar to those  observed  on  adherent
surfaces were evidenced  in  the  present  experiments.  Moreover,  confocal
imaging confirmed the presence of a pluricellular actomyosin  cable  at  the
edge of the closing open area (Figure 4). The contractility  of  this  cable
was  tested  with  two-photon  laser  photo-ablation  experiments   and   by
inhibiting myosin II with blebbistatin. When severed,  the  cable  retracted
within a few tens of seconds (Figure 4B, C, Supp. Video 2), indicating  that
it is under mechanical tension. By contrast, when  the  epithelium  bridging
over the non-adherent surface was punctured after closure, the hole did  not
expand upon ablation, indicating that no significant tension  is  stored  in
the monolayer itself. These small wounds then closed rapidly  by  developing
protrusions presumably on the debris left by the ablation. Furthermore,  the
addition of blebbistatin almost completely inhibited closure  (Supp.  Figure
4), whereas the same conditions have been shown to slow down  but  not  halt
closure on homogeneous adherent surfaces (4).

Our observations confirm that,  as  the  cells  do  not  interact  with  the
surface in our experiments, the contractile pluricellular  actomyosin  cable
along the edge must contract and pull  the  tissue  over  the  adhesion-free
surface by a  purse-string  mechanism.  By  contrast,  the  tension  in  the
epithelium itself is not a factor in this process.

To describe these experiments, we wrote the force balance  equation  at  the
free edge, on a line element of the contractile cable  of  radius  $r(t)$.  As
ingredients of the equation, we considered (Supp.  Figure  5):  1/  a  force
$f_{\mathrm{cable}} = -\gamma/r$ due to the line tension $\gamma$ of the 
contractile cable,  similar  to  what
has been proposed to describe the shape of  single  cells  anchored  to  the
surface via discrete points (28, 29); 2/ a friction force 
$f_{\mathrm{friction}} = -\xi \, \mathrm{d}r/\mathrm{d}t$ where  $\xi$
is a friction coefficient encapsulating the  dissipative  processes  at  the
cable and between cells (there is no interaction and hence  no  friction  at
the monolayer/substrate interface); and 3/ a stochastic force  
$f_{\mathrm{noise}}$  needed
to  model  the  above-described  stochastic  effects,  such  as   the   wide
distributions of closing times (Figure 2C-G, Supp. Figure  3)  or  the  very
noisy trajectories (Figure 2A). As puncturing the epithelium did not  result
in opening of the wound, we initially did not include epithelial tension  in
our description (see below). After dividing the force  balance  equation  by
the friction coefficient $\xi$, the Langevin equation describing  the  evolution
of the radius $r(t)$ reads (Supp. Note 1) (30):
\begin{equation}
  \label{eq:1}
  \frac{\mathrm{d}r}{\mathrm{d}t} = -\frac{\gre}{r} + \sqrt{2 \ds} \, \eta(t)
\end{equation}
where $\gre = \gamma/\xi$ and $\sqrt{2 \ds} \, \eta(t)$ is a noise term  
where  the  diffusion  coefficient  $\ds$
quantifies the amplitude of radius fluctuations at  the  margin.  Note  that
the  hypothesis  of  radial  force  balance  is  supported  by   independent
force measurements on keratinocytes, showing that, close to  the
free edge, the orthoradial component of the traction stress  remained  small
compared to the radial component during closure (17).

For the sake of simplicity, we further assumed that i/ $\eta(t)$  is  a  
Gaussian white noise with an autocorrelation function 
$\langle \eta(t) \eta(t')\rangle = \delta(t-t')$ and that ii/ $\gamma$, 
$\xi$   and $\ds$ remained  constant  (independent  of  $r$  and  $t$).  
Note  that  a  constant diffusion coefficient $\ds$ corresponds to 
fluctuations $\Delta \sigma$ of the  epithelial
tension $\sigma$ about its average value (zero in the  present  case): 
$\ds = \Delta \sigma^2/(2 \xi^2)$  (see below).

During  the  early  stages  of  fusion,  equation  (1)  reduces  to   simple
diffusion, and the initial mean square deviation reads (Supp. Note 4):
\begin{equation}
  \label{eq:2}
  \langle (r(0) - r(t))^2 \rangle_{t \to 0} = 2 D t
\end{equation}
The  experimental  data  were  in  good  agreement  with  this   theoretical
expression, yielding $\ds = 1.56 \pm 0.03 \, \mu \mathrm{m}^2 \mathrm{h}^{-1}$ 
(Figure 2I)  and  confirming  a
diffusive behavior of the radius at short times. Hence, the cable tension  
$\gamma$ does not contribute to the initial statistics that are fully  
determined  by the fluctuations.

In this framework, the fraction of closed wounds $f(R,t)$  obeyed  a  backward
Fokker-Planck equation (30) (see Supp. Notes) :
\begin{equation}
  \label{eq:3}
\DP{}{t}f(\ri,t) = - \frac{\gre}{R}\,\DP{}{\ri}f(\ri,t)
+D\,\DPn{2}{}{\ri}f(\ri,t)
\end{equation}
which could be solved numerically for a given set  of  parameters  and  with
boundary  conditions  in  accord  with  our  experimental  observations:  We
imposed that $r = R$ was a reflecting boundary (a  hole  never  opened  beyond
the area of the non-adhesive domain) and $r = 0$ was absorbing (there  was  no
re-opening after full closure). The closure time was then the time at  which
$r = 0$ was first attained (first-exit time).

A least-squares method allowed us to fit the model  to  the  data  over  the
whole map of the fraction of closed wounds (31). Varying $R$ and  $t$  at  
given $\gre$ and $\ds$,  we  minimized  the  mean  square  standardized  
error  between theoretical frequencies and experimental fractions 
(see  Supp.  Note  3  for the definition of the error function and 
a full description of  the  fitting procedure). This fit yielded  the  
following  estimates  of  the  parameters (Figure 3A):
\[
\gre = 10\;\mu\mathrm{m^{2} h^{-1}} \quad \mathrm{and}
\quad D = 1.6\;\mu\mathrm{m^{2} h^{-1}}
\]
where the numbers between brackets give the $95\, \%$ confidence interval  
(Figure 3C). Note that the diffusion coefficient is  consistent  with  
our  previous estimate based on the short-time evolution of the closure 
(Figure 2I).  With
these parameters, the  simulated  and  experimental  fractions  seemed  very
similar as can be seen in Figure 3A,B. More quantitatively,  the  particular
case of the closure half-time at which $50 \, \%$ of the domains  have  closed  as
well as the distributions of closure times for  various  radii  were  indeed
well-described by this set of parameters  (Figure  3D,  Figure  2C-G,  Supp.
Note 3). Finally, trajectories simulated from equation  (1)  (32)  with  the
previously  determined  values  of  $\gre$  and  $\ds$  closely   resembled   the
experimental ones (Figure 2A,B).

Altogether,  we  conclude  that  our  stochastic  model  provides  a   self-
consistent description of the closure dynamics. As the confidence  intervals
for $\ds$ and $\gre$ exclude $0$, the description is  also  minimal  in  the  sense
that none of the components can be removed from the description.

\section*{Discussion}

We have provided evidence that  a  cell  monolayer  can  develop  over  non-
adherent surfaces even when  cells  at  the  edge  are  not  anchored  to  a
substrate to pull on it. This property is the transposition  at  the  tissue
scale of a single cell's ability to bridge over defects smaller  than  their
own size. Experiments conducted in low calcium conditions or in presence  of
Blebbistatin show  that  such  closures  are  the  result  of  a  collective
behavior and that impairing the acto-myosin contractility, in particular  at
the purse-string cable, impairs this process.

The experimental results are well  described  by  a  stochastic  model  that
includes  the  tension  $\gamma$  of  the  circumferential  actomyosin  cable,   an
effective friction $\xi$ and the amplitude $\ds$  of  the  fluctuations  of  the
radius reflecting the ones of the epithelial tension. Our  theory  therefore
emphasizes the role and function of  the  purse-string  contractility  under
these conditions. Interestingly, this purse-string  mechanism  is  secondary
to lamellipodial protrusions when the  same  MDCK  cells  are  migrating  on
surfaces on which they can develop adhesions (3). As MDCK  cells  develop  a
peripheral actomyosin cable in both these situations, we conclude  that  the
nature of the substrate on which cells migrate controls whether  this  cable
has a regularization or a purse-string function.

From a typical value of the cable tension  $\gamma = 1-10$ nN  (25,  33),  
the  order  of
magnitude of the (one-dimensional) friction coefficient  is  
$\xi = 0.1-1 \, \mathrm{nN} \mu\mathrm{m}^{-2} \mathrm{h}$.  We  can
compare  this  value  with   the   hydrodynamic   two-dimensional   friction
coefficient measured for the same cells in  a  similar  setting  but  on  an
adherent substrate  (3):
$\xi_{\mathrm{adh}} =10^{-3} \, \mathrm{nN} \mu\mathrm{m}^{-3} \mathrm{h}$. 
The characteristic length defined  as $\xi/\xi_{\mathrm{adh}}$
is  typically  $100-1000 \, \mu$m,  consistent   with   the   correlation   length
characterizing the collective migration of MDCK  cells  on  glass  (3,  22).
Therefore, it is likely that the friction term originates  mostly  from  the
monolayer adhering to the glass around the non-adhesive domains.

An important (and intuitively unexpected) conclusion of our  study  is  that
fluctuations actively contribute to closing. This is  particularly  apparent
at the onset of closing (short times) where fluctuations  are  actually  the
dominant term in the closure dynamics (equation  (2)).  Theoretical  average
closure times can be  analytically  computed  as  first-exit  times  and  we
obtain for the average closure  time  $\langle t_c(R)\rangle = 
R^2/2(\gre+\ds)$  (Supp.  Note  3),  which  shows
immediately that, in a statistical sense, a non-zero  diffusion  coefficient
accelerates the closure.  The  model  further  predicts  that  the  standard
deviation of the closure time is proportional to and of same  order  as  the
average value (Supplementary notes 3,4). Quantitatively, within our  limited
dynamical range, these predictions are borne out by  data  with  the  fitted
parameters determined previously (Figure 2F,G). Last,  we  validate  one  of
our hypotheses of a white noise  in  the  diffusive  term  of  the  Langevin
equation (1). We used the model, and the fitted parameters, to  measure  the
experimental noise, and checked that the autocorrelation  function  of  this
noise decays rapidly with time, with a correlation time of the order  of  an
hour (Supplementary Figure 6). Since experiments are  performed  over  days,
this confirms that the white noise approximation is indeed appropriate.

We then  checked  whether  our  initial  assumption  of  not  including  the
epithelial tension in equation (1), initially based on tissue  photoablation
experiments, could be further confirmed. A non-zero tension $\sigma$  would  
add  a constant term $\sre = \sigma/\xi$ to the drift coefficient 
on the right-hand side  of  the
Langevin equation (1). As shown in Supp. Note 2,  analyzing  our  data  with
these three parameters $(\gre,\sre,\ds)$ 
yielded  a  very  small  (negative)  epithelial
tension whereas the two other parameters retained values close to  the  ones
previously  determined  (Supp.  Figure  7A,B).  Therefore,  the   epithelial
tension can be safely omitted, suggesting that the time-scales  involved  in
the fusion process are sufficiently long to allow elastic stresses to  relax
(34). Not surprisingly, putting $\gre = 0$ in this new equation led to a  drastic
decrease of the fraction of closed wounds at $83$ h  in  good  agreement  with
our experimental observations in presence of Blebbistatin (Supp. Figure  4).

Next, we asked whether including fluctuations of  the  cable  tension  would
better describe our experimental data than our current hypothesis  in  which
these fluctuations arise only from the epithelial tension ($\ds$ independent  of
$r$). Assuming for simplicity that fluctuations in the monolayer  tension  and
in the  cable  tension  are  not  correlated,  we  expressed  the  diffusion
coefficient as $D_{\mathrm{tot}}(r) = \ds + \dg/r^2$ where $\ds$ and  
$\dg$ are  proportional  to  the  (constant)
amplitudes of the fluctuations of these two tensions. Fitting model to  data
within a three-dimensional parameter space  $(\gre,\ds,\dg)$  
allowed  us  to  conclude
that fluctuations in the cable tension could be ignored  except  for  values
of $r$ much smaller than a cell size (Supp. Notes 3 and  Supp.  Figure  7C,D).
Of note, at these very small radius values ($r < 1 \, \mu$m), the  fluctuations  of
the cable become dominant over  both  the  fluctuations  of  the  epithelial
tension and the deterministic cable tension  (Supp.  Note  2).  This  regime
corresponds to the very late stages of closure,  which  are,  unfortunately,
beyond our experimental time resolution. Apart from this regime,  the  noise
term of the stochastic model therefore originates from fluctuations  of  the
monolayer tension about its zero mean value. These fluctuations result  from
cell-level  dynamics   occurring   over   short   time   scales,   such   as
rearrangements, divisions, or cells  being  pulled  out  from  the  adhesive
part.

From the typical value of $\xi$ determined  above,  we  can  also  estimate  the
magnitude of the active fluctuations of the epithelial tension: 
$\Delta\sigma^2_{\mathrm{active}} = 2 \ds \xi^2 = 10^{-2} - 1 \,
\mathrm{nN}^2 \, \mu\mathrm{m}^{-2} \, \mathrm{h}$.  This
is, to our knowledge, the first experimental measurement of  this  quantity.
We  can  compare  this  value  to  the  amplitude  of  tension  fluctuations
resulting from thermal noise: $\Delta \sigma^2_{\mathrm{thermal}} = 
2 D_{\mathrm{thermal}} \xi^2$. As the total  friction  of  the  cable
over its perimeter is $2 \pi R \xi$, the amplitude of  thermal  radius  fluctuations
is $D_{\mathrm{thermal}} = k_B T/(2 \pi R \xi)$ where $k_B$ is Boltzmann's constant, $T \sim 300$ K and $R \sim 10 \, \mu$m.  Hence,
$\Delta \sigma^2_{\mathrm{thermal}} = 10^{-8} - 10^{-7} \,
\mathrm{nN}^2 \, \mu\mathrm{m}^{-2} \, \mathrm{h}$ : the amplitude of  the  epithelial  tension  fluctuations  is  several
orders of magnitude larger than that of the  thermal  tension  fluctuations,
as expected in these active cellular systems (35).

We end  by  considering  tissue  fusion  over  longer  periods  of  time.  A
prediction of our model is that arbitrarily large wounds will always  close,
given an arbitrarily large duration, as the stochastic  process  defined  by
equation (1) yields trajectories that eventually  always  reach  zero  (36).
Practically, however, because of the change of behavior induced  by  the  3D
rim that develops at the  adherent/non-adherent  boundary,  we  stopped  our
analysis at $83$ h. Several arrays were  further  analyzed  over  longer  time
periods: at $4$ days, the evolution of the  fractions  of  closed  wounds  was
still well described by our model (Figure 2H). However, the closure  process
stopped after that time as we observed no evolution between  $4$  and  $7$  days
(Supp. Figure 8), presumably due to rim formation at the  edge  even  though
the cells were  still  active  and  the  border  of  the  wounds  was  still
fluctuating. This rim is likely to be the reason  why  large  wound  do  not
close in contrast with other cell types such as keratinocytes that are  able
to close gaps of several $100 \,\mu$m (17).

In conclusion, we have shown that the collective migration of an  epithelium
can switch between two modes, depending on the  cells' affinity  for  their
substrate. Whereas on adhesive surfaces, the collective migration is  mostly
driven by protrusions, our work shows that  the  purse-string  mechanism  is
essential on non-adherent surfaces. Importantly, the active fluctuations  of
the tissue are also crucial and accelerate the closure dynamics.  A  natural
future extension of our work will be to  elaborate  a  more  complete  model
where these active fluctuations are combined with the viscoelastic  rheology
of the tissue (17), to get a full description of the  closure  processes  in
more complex situations  including,   for
example,  in  vivo  tissue  fusion  in  embryonic  morphogenesis,   or   the
collective migration of cancer cells in fibrillar environments  (37).  Other
cell models may be better suited to address  these  important  issues  where
gaps larger than the ones studied in the present work are to be closed.

\begin{figure}[!th]
\centering
\includegraphics[scale=0.6]{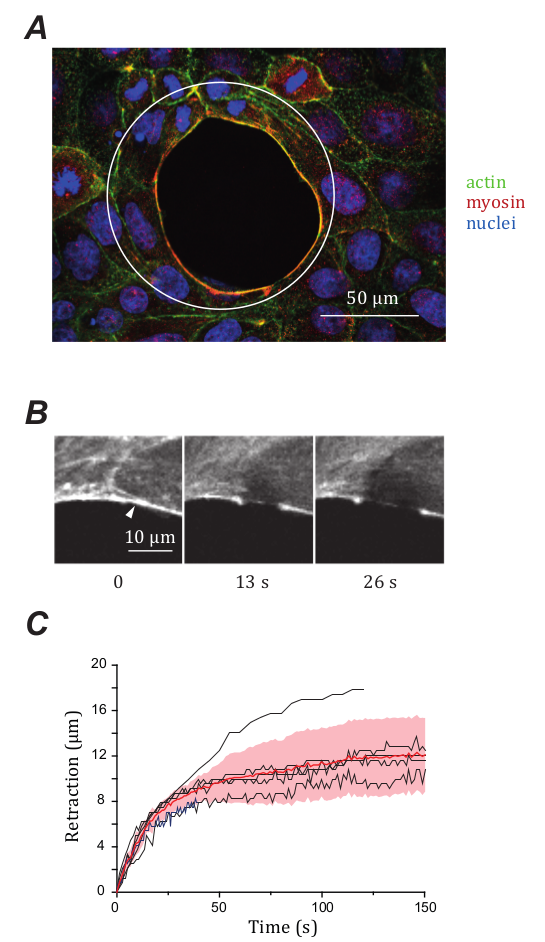}
\caption{
\textbf{A contractile pluricellular actomyosin cable localizes at the
border of the monolayer.}
\textbf{A} Colocalisation of actin and myosin at the free
edge of the closing epithelium. The white circle is the underlying domain.
\textbf{B} Ablation of the cable (ablation point is figured by the white 
triangle) at t=0 and its subsequent retraction.
\textbf{C} Dynamics of retraction of cables
on several wounds. The red line is the average retraction, the pink area is
the standard deviation.
}
\end{figure}

\section*{Methods}

\subsection*{Preparation of the patterns}

Non-adherent domains of  varying  radii  were  micropatterned  onto  cleaned
glass cover slips. The glass  slides  were  first  uniformly  treated  by  a
surface treatment of polyacrylamide and polyethyleneglycol  (PEG)  on  which
cells do not adhere (18, 19, 38). Domains whose radii were between $5$ and  
$75 \,\mu$m with a $0.5 \,\mu$m increment were defined by photolithography 
directly on  the coating in such a way that it remained protected by the 
photoresist  (S1813, Microchem, Westborough, MA) at the  desired  location  
of  the  non-adhesive domains (19). Using the photoresist as an etching mask, 
the PEG coating  was removed in the photoresist-free areas with an  air  
plasma  (Harrick  plasma cleaner), revealing the underlying glass.  
The  resist  was  then  dissolved leading to PEG-coated domains surrounded  
by  a  clean  glass  surface.  The surface treatment was stable for weeks 
in biological buffers (19).

\subsection*{Cell culture}

MDCK  cells  (39)  were  cultured  in  Dulbecco's modified  Eagle's  medium
supplemented with $10 \, \%$ FBS (Sigma), $2$ mM L-glutamin solution 
(Gibco)  and  $1 \, \%$ antibiotic  solution  (penicillin  
($10,000$   units/mL),   streptomycin   ($10$ mg/mL)). Cells were seeded 
and maintained at $37^\circ$C, $5 \, \%$CO2  and  $90 \, \%$  humidity
throughout the  experiments.  We  also  used  MDCK  LifeAct  cells  (3)  for
ablation experiments (these clones were cultured in  presence  of  geneticin
($400 \, \mu$g/mL)).

Blebbistatin (Sigma) was used at a concentration of $50\,\mu$M. 
Experiments  were started in the absence of  the  drug.  At  confluence,  
a  fraction  of  the supernatant was pumped out, mixed with the drug  and  
re-injected  into  the well.

Low-calcium medium (calcium-free DMEM, fetal bovine serum $10 \, \%$, 
Penstrep  $1 \, \%$, $50$ mM calcium) was used  to  reduce  cell-cell  
adhesion.  Experiments  were started in regular DMEM and the buffer 
was changed to  low-calcium  DMEM  at confluence.

\subsection*{Microscopy and data analysis}

The bottoms of Petri dishes or 6-well plates were  replaced  with  patterned
glass slides. Cells were  imaged  in  phase  contrast  on  an  Olympus  IX71
inverted microscope equipped with temperature, CO2 and  humidity  regulation
(LIS), a motorized stage for multipositioning (Prior)  and  a  Retiga  4000R
camera (QImaging). Unless otherwise specified, a $10 \times$ objective was used  and
images were acquired every $30$ min. Displacements and image acquisition  were
computer-controlled with Metamorph (Molecular Devices).

Fixed fluorescently marked cells were observed under an  upright  Imager  Z2
spinning disk  microscope  (Zeiss,  Oberkochen,  Germany)  equipped  with  a
CoolSnapHQ2 camera (Photometrics, Tuscon, AZ)  and  a  $63 \times$  water  immersion
objective.  All  acquisitions  were  controlled  using  MetaMorph   software
(Molecular Devices, Sunnyvale, CA).

Images  were  processed  with  the  ImageJ  software  
or  with  Matlab
(MathWorks,  Natick,  MA)  routines.  Further  analysis   was   occasionally
performed on Origin (OriginLab, Northampton, MA).

Unless otherwise specified,  fractions  were  computed  from  at  least  100
domains for each size measured over at least 4 distinct experiments.

\subsection*{Laser ablations}

Photoablation  experiments  were  performed  on  an  LSM  710  NLO   (Zeiss)
microscope equipped with a two-photon MaiTai laser and a $40 \times$  oil  immersion
objective. The two-photon laser was used at $85 \,\%$ power and  at  
a  wavelength of $890$ nm.

\subsection*{Immunoflurocescence}

Cells were fixed in $4 \,\%$ PFA, permeabilized in $0.1 \,\%$ 
Triton X-100  and  blocked in $10 \,\%$ FBS in PBS. 
Vinculin labeling was performed with a  mouse  monoclonal
anti-vinculin antibody (Sigma, 1:500) and Paxillin  labeling  was  performed
with a mouse anti-paxillin antibody (Sigma, 1:500) both  followed  by  Alexa
488 donkey anti-mouse (Life Technologies, 1:500). Actin  was  labeled  using
Alexa 546 phalloidin (Life Technologies, 1:1000). Myosin  was  labeled  with
rabbit anti-phospho myosin light chain (Ozyme, Saint  Quentin  en  Yvelines,
France,  1:100)  followed   by   Alexa   488   chicken   anti-rabbit   (Life
Technologies, 1:1000). Hoescht 33342 (Sigma, 1:10000) was used to  mark  the
nuclei.

\section*{Acknowledgements}

We gratefully thank Isabelle Bonnet, Axel Buguin, Nir Gov, Jonas  Ranft  and
all the members of the ``biology-inspired physics at  mesoscales'' group  for
discussions, as well as Gabriel Dumy for performing part of the analysis.
HGY thanks the Fondation Pierre-Gilles de Gennes for financial  support.  We
acknowledge financial support from the  Programme  Incitatif  et  Coop\'eratif
Curie ``Mod\`eles Cellulaires''.
The ``biology  inspired  physics  at  mesoscales''  group  and  the  ``physical
approaches of biological  problems''  group  are  part  of  the  CelTisPhyBio
Labex.
We acknowledge the Cell and  Tissue  Imaging  Platform  (member  of  France-
Bioimaging)  of  the   Genetics   and   Developmental   Biology   Department
(UMR3215/U934) of Institut  Curie  and  in  particular  Olivier  Renaud  and
Olivier Leroy.

\section*{References}

1.    Ray HJ, Niswander L (2012) Mechanisms of tissue fusion during
development. Development 139:1701-11.

\noindent
2.    Martin P (1997) Wound healing--aiming for perfect skin regeneration.
Science 276:75.

\noindent
3.    Cochet-Escartin O, Ranft J, Silberzan P, Marcq P (2014) Border forces
and friction control epithelial closure dynamics. Biophys J 106:65-73.

\noindent
4.    Anon E et al. (2012) Cell crawling mediates collective cell migration
to close undamaged epithelial gaps. Proc Natl Acad Sci U S A 109:10891-6.

\noindent
5.    Brugu\'es A et al. (2014) Forces driving epithelial wound healing. Nat
Phys 10:683.

\noindent
6.    Reffay M et al. (2014) Interplay of RhoA and mechanical forces in
collective cell migration driven by leader cells. Nat Cell Biol 16:217.

\noindent
7.    Abreu-Blanco MT, Verboon JM, Liu R, Watts JJ, Parkhurst SM (2012)
Drosophila embryos close epithelial wounds using a combination of cellular
protrusions and an actomyosin purse string. J Cell Sci 125:5984-97.

\noindent
8.    Copp AJ, Brook FA, Estibeiro JP, Shum AS, Cockroft DL (1990) The
embryonic development of mammalian neural tube defects. Prog Neurobiol
35:363-403.

\noindent
9.    Bement WM, Mandato C, Kirsch MN (1999) Wound-induced assembly and
closure of an actomyosin purse string in Xenopus oocytes. Curr Biol
9:579-87.

\noindent
10.   Kiehart DP (1999) Wound healing: The power of the purse string. Curr
Biol 9:R602-605.

\noindent
11.   Jacinto A, Martinez-Arias A, Martin P (2001) Mechanisms of epithelial
fusion and repair. Nat Cell Biol 3:E117-E123.

\noindent
12.   Hutson MS et al. (2003) Forces for morphogenesis investigated with
laser microsurgery and quantitative modeling. Science 300:145-9.

\noindent
13.   Friedl P, Locker J, Sahai E, Segall JE (2012) Classifying collective
cancer cell invasion. Nat Cell Biol 14:777-783.

\noindent
14.   Curtis A, Varde M (1964) Control of cell behavior: topological
factors. J Natl Cancer Inst 33:15-26.

\noindent
15.   Vedula SRK et al. (2014) Epithelial bridges maintain tissue integrity
during collective cell migration. Nat Mater 13:87-96.

\noindent
16.   Gautrot JE et al. (2012) Mimicking normal tissue architecture and
perturbation in cancer with engineered micro-epidermis. Biomaterials
33:5221-9.

\noindent
17.   Vedula SRK et al. (2015) Mechanics of epithelial closure over non-
adherent environments. Nat Commun 6:6111.

\noindent
18.   Tourovskaia A, Figueroa-Masot X, Folch A (2006) Long-term
microfluidic cultures of myotube microarrays for high-throughput focal
stimulation. Nat Protoc 1:1092-104.

\noindent
19.   Deforet M, Hakim V, Yevick HG, Duclos G, Silberzan P (2014) Emergence
of collective modes and tri-dimensional structures from epithelial
confinement. Nat Commun 5:3747.

\noindent
20.   Underhill GH, Galie P, Chen CS, Bhatia SN (2012) Bioengineering
methods for analysis of cells in vitro. Annu Rev Cell Dev Biol 28:385-410.

\noindent
21.   Harris AR et al. (2012) Characterizing the mechanics of cultured cell
monolayers. Proc Natl Acad Sci U S A 109:16449-54.

\noindent
22.   Petitjean L et al. (2010) Velocity fields in a collectively migrating
epithelium. Biophys J 98:1790-800.

\noindent
23.   Geiger B, Spatz JP, Bershadsky AD (2009) Environmental sensing
through focal adhesions. Nat Rev Mol Cell Biol 10:21-33.

\noindent
24.   Bischofs IB, Safran S, Schwarz US (2004) Elastic interactions of
active cells with soft materials. Phys Rev E 69:1-17.

\noindent
25.   Guthardt Torres P, Bischofs IB, Schwarz US (2012) Contractile network
models for adherent cells. Phys Rev E 85:1-13.

\noindent
26.   Rossier OM et al. (2010) Force generated by actomyosin contraction
builds bridges between adhesive contacts. EMBO J 29:1055-68.

\noindent
27.   Kim JH et al. (2013) Propulsion and navigation within the advancing
monolayer sheet. Nat Mater 12:856-863.

\noindent
28.   Bar-Ziv R, Tlusty T, Moses E, Safran S, Bershadsky AD (1999)
Pearling in cells: A clue to understanding cell shape. Proc Natl Acad Sci
U S A 96:10140-10145.

\noindent
29.   Bischofs IB, Klein F, Lehnert D, Bastmeyer M, Schwarz US (2008)
Filamentous network mechanics and active contractility determine cell and
tissue shape. Biophys J 95:3488-96.

\noindent
30.   Gardiner WC (2004) Handbook of stochastic methods for physics,
chemistry and the natural sciences (Springer-Verlag, Berlin).

\noindent
31.   Bevington PR, Robinson DK (1969) Data reduction and error analysis
for the physical sciences (McGraw-Hill, New York).

\noindent
32.   Kloedel P, Platen E (1999) Numerical simulations of stochastic
differential equations (Springer).

\noindent
33.   Yoshinaga N, Marcq P (2012) Contraction of cross-linked actomyosin
bundles. Phys Biol 9:046004.

\noindent
34.   Ranft J et al. (2010) Fluidization of tissues by cell division and
apoptosis. Proc Natl Acad Sci U S A 107:20863-8.

\noindent
35.   Douezan S, Brochard-Wyart F (2012) Active diffusion-limited
aggregation of cells. Soft Matter 8:784.

\noindent
36.   Martin E, Behn U, Germano G (2011) First-passage and first-exit times
of a Bessel-like stochastic process. Phys Rev E 83:051115.

\noindent
37.   Yevick HG, Duclos G, Bonnet I, Silberzan P (2015) Architecture and
migration of an epithelium on a cylindrical wire. Proc Natl Acad Sci
112:5944-5949.

\noindent
38.   Tourovskaia A et al. (2003) Micropatterns of Chemisorbed Cell
Adhesion-Repellent Films Using Oxygen Plasma Etching and Elastomeric Masks.
Langmuir 19:4754-4764.

\noindent
39.   Bellusci S, Moens G, Thiery J-P, Jouanneau J (1994) A scatter factor-
like factor is produced by a metastatic variant of a rat bladder carcinoma
cell line. J Cell Sci 107:1277-87.


\FloatBarrier

\newpage 
\setcounter{figure}{0}

\begin{figure}[ht]
\centering 
\includegraphics[width=0.9\textwidth]{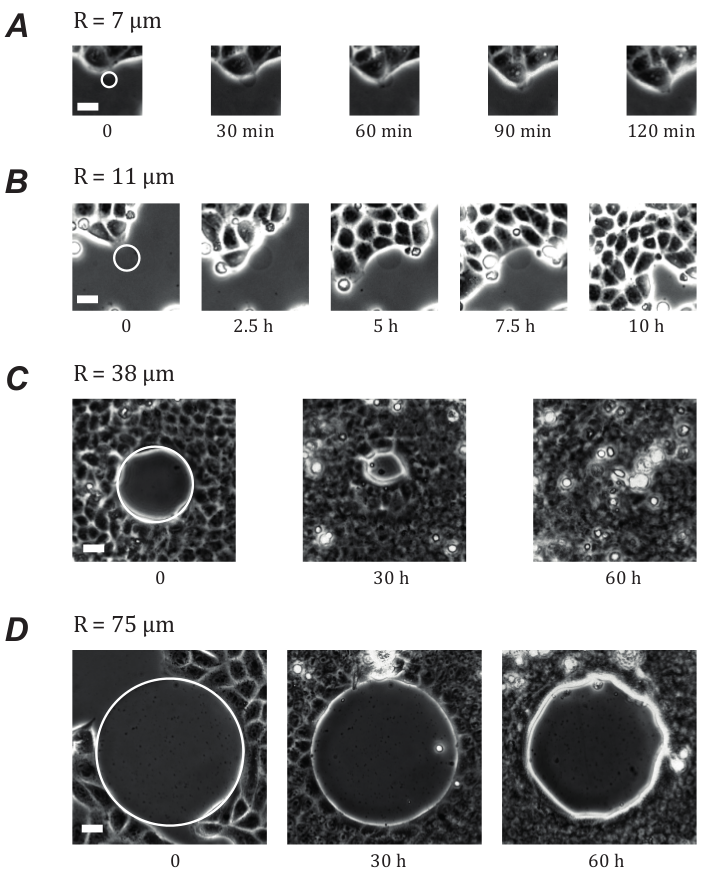}
\captionsetup{labelformat=support}
\caption{
\textbf{Behaviors of the monolayer according to the radius
of the non-adherent domain.}
\textbf{A} Domains smaller than a cell size are covered
rapidly with no arrest of the monolayer.
\textbf{B} On domains whose size is
comparable with a cell size, the monolayer stops before covering them
rapidly as it progresses.
\textbf{C} For still larger domains ($30 \, \mu \mathrm{m} < R < 65 \, \mu$m),
the monolayer surrounds the domains and then covers them by purse string as
described in the text.
\textbf{D} For domains larger than $70 \, \mu$m, the monolayer
surrounds the domains but cannot cover them.
}
\end{figure}

\begin{figure}[ht]
\centering 
\includegraphics[width=0.9\textwidth]{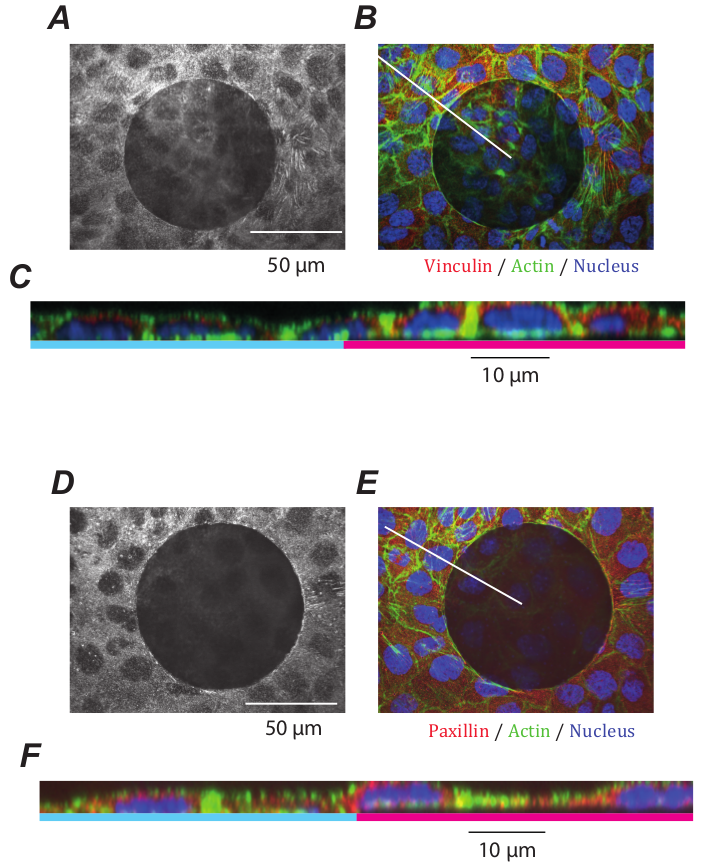}
\captionsetup{labelformat=support}
\caption{
\textbf{The cells do not develop adhesions with their substrate.}
\textbf{A-B} and \textbf{D-E} show no significant signal of the 
adhesion proteins vinculin (A) or paxillin (D) at the basal plane. 
\textbf{C,F} On the $xz$ sections a
thin line void of proteins can be seen over the non-adherent surface
(figured in red while the adhering surface is blue). The white lines in
panels B and D are the plane of the sections C and F. 
A-C: $R = 41 \, \mu$m; D-F: $R = 45.5 \, \mu$m.
}
\end{figure}

\begin{figure}[ht]
\centering 
\includegraphics[width=0.5\textwidth]{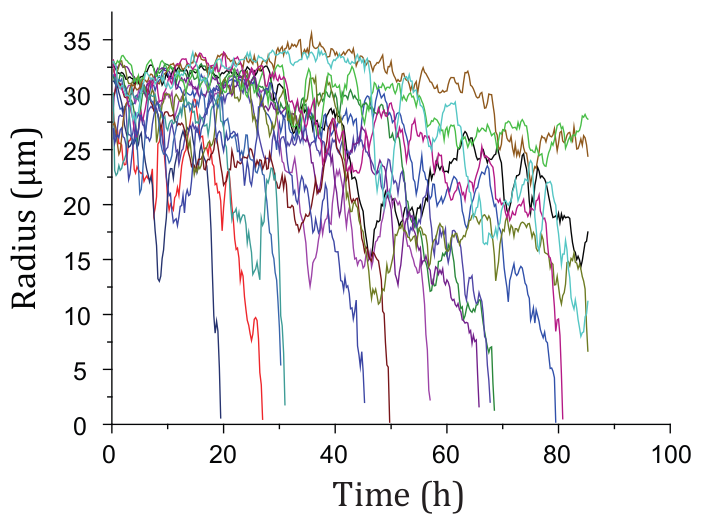}
\captionsetup{labelformat=support}
\caption{
\textbf{Variability of the trajectories.} Individual
trajectories are noisy but also define a very broad distribution of closure
times. $R = 35\,\mu$m.
}
\end{figure}

\begin{figure}[ht]
\centering 
\includegraphics[width=0.5\textwidth]{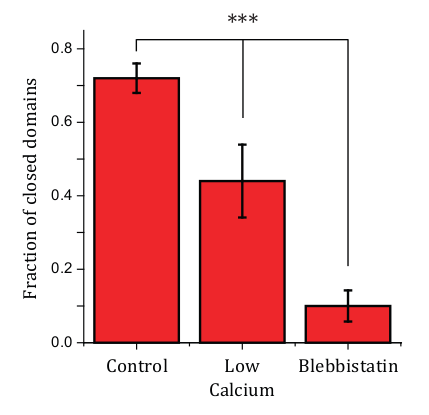}
\captionsetup{labelformat=support}
\caption{
\textbf{Importance of the monolayer cohesion and contractility.}
Low calcium conditions that lead to less cohesive monolayers
resulted in a lower fraction of closed wounds as did the addition of
Blebbistatin that almost halted the closure ($N_{\mathrm{control}}=125$, 
$N_{\mathrm{low}\,\mathrm{calcium}}=25$, $N_{\mathrm{blebbistatin}}=50$). 
$R = 40 \,\mu$m. Error bars are standard deviations.
}
\end{figure}

\begin{figure}[ht]
\centering 
\includegraphics[width=0.5\textwidth]{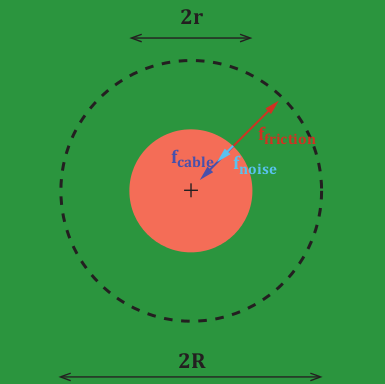}
\captionsetup{labelformat=support}
\caption{
\textbf{Schematics of the model.} Force balance of a line
element at the wound margin, for a circular wound of radius $r$. The three
lineic force densities depicted are: the tension of the actomyosin cable, a
viscous friction force, and a fluctuating force. The initial radius of the
non-adhesive patch is $R$. (the cell monolayer is represented in green; the
non-adherent surface in red).
}
\end{figure}

\begin{figure}[ht]
\centering 
\includegraphics[width=0.5\textwidth]{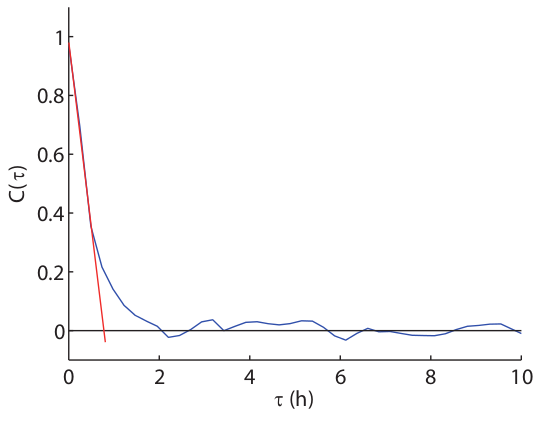}
\captionsetup{labelformat=support}
\caption{
\textbf{Noise autocorrelation function.}
 We use the Langevin equation (1) to measure the noise term  
$\eta(t) = \frac{1}{\sqrt{2D}} \, \left( \dot{r} + \frac{\gre}{r}\right)$
for each trajectory, with the fitted parameter values 
$\gre = 10\;\mu\mathrm{m^{2} h^{-1}}$ and  $D = 1.6\;\mu\mathrm{m^{2} h^{-1}}$.  
The noise autocorrelation function is obtained by ensemble averaging: 
$C(\tau) = \langle \eta(t) \eta(t+\tau) \rangle$. 
It decays quickly, with a correlation time of the order of $1$h.
}
\end{figure}

\begin{figure}[ht]
\centering 
\includegraphics[width=0.9\textwidth]{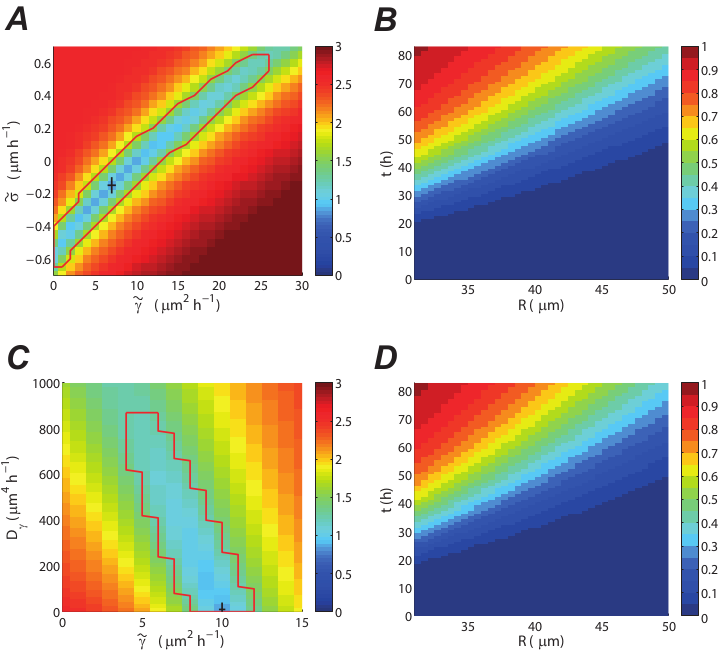}
\captionsetup{labelformat=support}
\caption{
\textbf{Model robustness.}
\textbf{A,B} Influence of an epithelial tension. (A) $(\gre,\sre)$ 
plane of the error landscape at $D = 1.5\;\mu\mathrm{m^{2} h^{-1}}$ 
(logarithmic scale). The cross denotes the optimum, the red contour bounds 
the confidence region. (B) Optimal closure
frequency map (linear scale), computed for 
$\gre = 7\;\mu\mathrm{m^{2} h^{-1}}$, $D = 1.5\;\mu\mathrm{m^{2} h^{-1}}$
and  $\sre = -0.15\;\mu\mathrm{m^{2} h^{-1}}$.
\textbf{C,D} Influence of fluctuations of the cable tension. 
(C) $(\gre,\dg)$ plane of the error landscape at 
$D = 1.5\;\mu\mathrm{m^{2} h^{-1}}$ (logarithmic scale). The cross denotes the
optimum, the red contour bounds the confidence region. (D) Optimal closure
frequency map (linear scale), computed for
$\gre = 10\;\mu\mathrm{m^{2} h^{-1}}$, $D = 1.5\;\mu\mathrm{m^{2} h^{-1}}$
$\dg = 10\;\mu\mathrm{m^{2} h^{-1}}$.  
}
\end{figure}

\begin{figure}[ht]
\centering 
\includegraphics[width=0.5\textwidth]{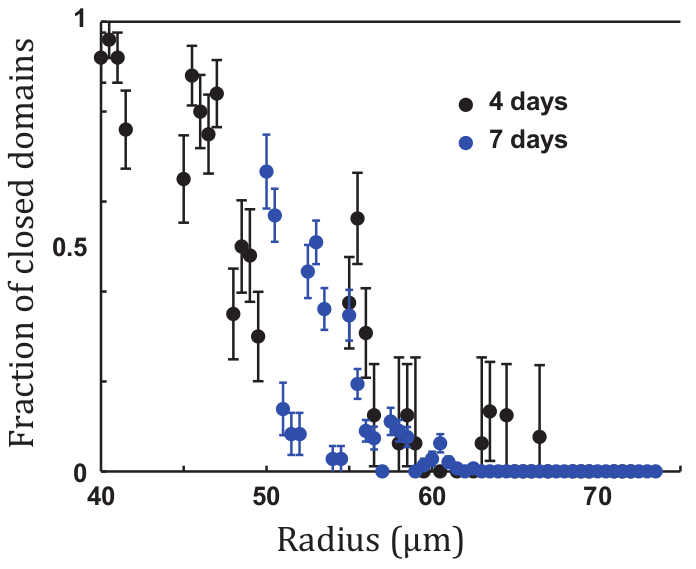}
\captionsetup{labelformat=support}
\caption{
\textbf{Long time behavior.}
Fractions of closed wounds at $4$
days (black points) and $7$ days (blue points). Because of the formation of a
peripheral rim, there is no evolution after 4 days. Error bars are SEMs.
}
\end{figure}

\FloatBarrier 
\centerline{\Large \textbf{Supplementary Notes}}

\bigskip
\section{A stochastic description}
\label{sec:model}

\subsection{A Langevin equation}
\label{sec:model:Langevin}

Wound closure dynamics on a nonadhesive patch is a noisy process, 
with, \emph{e.g.}, large fluctuations of the closure time
$t_{c}$ at a given radius $\ri$.
This observation calls for a stochastic description. Since the wound is
approximately invariant by rotation about its center, 
force balance on a line element
at the margin may be expressed as a stochastic
equation for the wound radius $\rt(t)$
\begin{equation}
-\friction\dot{\rt}+F_{1}(\rt,t)+F_{2}(\rt,t)=0\label{eq:Langevin:v0}
\end{equation}
 where $F_{1}(\rt,t)$ and $F_{2}(\rt,t)$ respectively denote the deterministic
and the stochastic component of the lineic force density, 
and $\friction$ is an effective 
friction coefficient subsuming all dissipative processes at play.
Since ablation experiments show that the circumferential actomyosin cable 
is under tension, we write 
\begin{equation}
F_{1}(\rt,t)=-\frac{\gnre}{\rt} 
\label{eq:def:F1}
\end{equation}
 where $\gnre$ denotes the line tension of the contractile cable 
(see Supp. Fig.~5).

A simple stochastic model of the dynamics may be written as 
\begin{equation}
\dot{\rt}(t)=D_{1}(\rt,t)+\sqrt{2D_{2}(\rt,t)} \, \eta(t)\label{eq:Langevin:v1}
\end{equation}
 where $D_{1}$ and $D_{2}$ respectively denote the drift and diffusion 
coefficients, and $\eta(t)$ is a Gaussian white noise with an autocorrelation
function $\langle\eta(t)\eta(t')\rangle=\delta(t-t')$. 
Dividing the parameter $\gnre$ 
by the friction coefficient $\friction$ gives a drift coefficient: 
\begin{equation}
D_{1}(\rt,t)=-\frac{\gre}{\rt},
\label{eq:def:D1}
\end{equation}
with $\gre=\gnre/\friction$.

Unless explicitly mentioned, 
we study in the following the stochastic dynamics generated
by the Langevin equation 
\begin{equation}
\dot{\rt}(t)=-\frac{\gre}{\rt}+  \sqrt{2D} \,\eta(t)
\label{eq:Langevin:v2}
\end{equation}
with a \emph{constant} diffusion coefficient $D$
\begin{equation}
D_{2}(\rt,t)= D
\label{eq:def:D2}
\end{equation}
We observe that collective migration always tends to close the cell-free space, 
which never opens up beyond the area of the non-adhesive patch. 
The boundary condition at $\rt=\ri$ is therefore \emph{reflecting}. 
Since closed wounds do not re-open, the boundary condition at $\rt=0$ 
is \emph{absorbing}.
In numerical simulations of the Langevin equation, 
using a finite time step $h$ [32], 
the boundary conditions are implemented as follows:
\begin{itemize}
\item[\emph{(i)}]{absorbing condition at $\rt = 0$:
the simulation stops whenever the radius becomes negative  $\rt(t+h)<0$;}
\item[\emph{(ii)}]{reflecting condition at $\rt = \ri$:
a radius larger than $\ri$,  $\rt(t+h)>\ri$ is replaced by 
a radius smaller than $\ri$, $2\ri-\rt(t+h) < \ri$.}
\end{itemize}

\subsection{A Fokker-Planck equation}
\label{sec:model:FP}

Eq.~(\ref{eq:Langevin:v1}) is equivalent to an evolution equation
for the transition probability distribution function 
$p(\rt,t\mid \ri,0)$ between a radius $\ri$ at the initial time $0$ 
and a radius $\rt$ at time $t$ [30].
It is convenient to write the backward Fokker-Planck
equation for $p(\rt,t\mid \ri,0)$:
\begin{equation}
\DP{}{t} p(\rt,t\mid \ri,0)=
D_{1}(\rt,t)\,\DP{}{\ri}p(\rt,t\mid \ri,0)+
D_{2}(\rt,t)\,\DPn{2}{}{\ri}p(\rt,t\mid \ri,0)
\label{eq:def:FP:1}
\end{equation}
The probability $f(\ri,t)$ that a patch with radius $\ri$ is closed 
at time $t$, or closure frequency, reads
\begin{equation}
  \label{eq:def:closure:pdf}
 f(\ri,t) = \mathrm{Prob}(t_{c}(\ri)\leq t)
=1 - \int_{0}^{\ri}p(\rt,t\mid \ri,0)\, \mathrm{d}\rt 
\end{equation}
Integrating (\ref{eq:def:FP:1}) over $\rt$, we obtain  the evolution equation
for the closure frequency 
\begin{equation}
\DP{}{t}f(\ri,t) = D_{1}(\ri,t)\,\DP{}{\ri}f(\ri,t)+D_{2}(\ri,t)\,\DPn{2}{}{\ri}f(\ri,t)
\label{eq:def:FP:2}
\end{equation}
Since wounds are initially open, the initial condition is 
$f(R_{\mathrm{i}},0)=0$ for an initial radius $\ri = R_{\mathrm{i}}$.
For all time $t$, we naturally impose  $f(0,t)=1$ at $\ri=0$, and 
the reflecting boundary condition at $R_{\mathrm{i}}$ reads
$\DP{f}{\ri}(R_{\mathrm{i}},t)=0$ [30].

Using this set of initial and boundary conditions, the numerical resolution 
of Eq.~(\ref{eq:def:FP:2}) is performed
with the function \texttt{pdepe} of Matlab. 
For a given set of model parameters, 
Eq.~(\ref{eq:def:FP:2}) is solved on the interval $R\in[0,R_{\mathrm{i}}]$.
The value at $R_{\mathrm{i}}$,  $\fsim(R_{\mathrm{i}},t) = f(\ri=R_{\mathrm{i}},t)$
can then be compared with the experimentally measured fraction
$\fexp(R_{\mathrm{i}},t)$ of wounds of initial radius $R_{\mathrm{i}}$ 
closed at time $t$.

From Eq.~(\ref{eq:def:closure:pdf}), $f(\ri,t)$ is also the
cumulated distribution function of closure times.
As a consequence, the distribution function $P(\tc \mid \ri)$ of the
closure times $\tc$ of patches of radius $\ri$ is obtained
by differentiating  $f$ with respect to time:
\begin{equation}
  \label{eq:def:pdf:tc}
  P(\tc \mid \ri) = \DP{f}{t}(\ri,\tc)
\end{equation}

\section{Parameter fitting and model selection}
\label{sec:fit}

\subsection{A least-squares method}
\label{sec:fit:leastsquares}

In this section, the theoretical closure frequencies 
$\fsim(\ri,t \mid \gre, D)$ are computed numerically
for a given set of parameter values $( \gre, D )$. 
Initial and boundary conditions for the  evolution equation
\begin{equation}
\DP{}{t}f(\ri,t)= - \frac{\gre}{\ri} \,\DP{}{\ri}f(\ri,t)+
D \,\DPn{2}{}{\ri}f(\ri,t)
\label{eq:def:FP:3}
\end{equation}
are given in Sec.~\ref{sec:model:FP}.
Varying $(R,t)$ at fixed $(\gre,D)$,
we calculate the mean square standardized error $E^2(\gre,D)$ between  
theoretical frequencies and experimental fractions [31]: 
\begin{equation}
E^{2}(\gre, D)= \frac{1}{N}\, \sum_{\ri,t}  
\frac{ \left(\fsim(\ri,t \mid \gre, D) - 
\fexp(\ri,t)\right)^{2}}{\sigma_{\fexp}^2(\ri,t)}
\label{eq:def:mse}
\end{equation}
where $N$ is the total number of data points, and
$\sigma_{\fexp}^2(\ri,t)$ is the variance of the fraction of patches 
of radius $R$ closed at time $t$, as measured
over $\nexp(\ri)$ experiments: 
\begin{equation}
  \label{eq:def:var:exp}
  \sigma_{\fexp}^2(\ri,t) = 
\frac{f_{exp}(\ri,t) \left(1-f_{exp}(\ri,t) \right)}{\nexp(\ri)}
\end{equation}
The experimental fractions $\fexp(\ri,t)$ are measured
from a total of $6625$ patches, with radii ranging from $31$ to
$49.5\,\mu$m with a $0.5\,\mu$m step, over a total duration $T=83$h,
and with a time resolution $\Delta t=1$h. In practice, 
we define  $\mathrm{N_{\ri}}=13$ bins of width $\Delta \ri=1.5\,\mu$m
and $\mathrm{N_{t}=84}$ time points per radius,
so that $N=N_\ri \, N_t = 1092$.

The error landscape is shown in Fig.~3C. 
The minimum of the mean square error is achieved for
\begin{eqnarray}
\gre_{\mathrm{min}} & = & 10\;\mu\mathrm{m^{2} h^{-1}}
\label{eq:best:gamma1}\\
D_{\mathrm{min}} & = & 1.6\;\mu\mathrm{m^{2} h^{-1}}
\label{eq:best:D}
\end{eqnarray}
corresponding to a minimal mean square error value 
\begin{equation}
  \label{eq:min:error}
  E^2_{\mathrm{min}} = \mathrm{min}_{\gre,D}\, E^2(\gre,D)=
  E^2(\gre = \gre_{\mathrm{min}}, D= D_{\mathrm{min}}) \simeq 7.6
\end{equation}
at the bottom of a well-defined single well.
The optimal parameter values  (\ref{eq:best:gamma1}-\ref{eq:best:D}) 
yield the best agreement with experimental data
(compare Figs.~3A and 3B).

At a given radius $R$ and time $t$,
the $95 \, \%$ confidence interval of the optimal theoretical frequency
$\fsim^{\mathrm{min}} = \fsim(\ri,t \mid \gre_{\mathrm{min}},  D_{\mathrm{min}})$ 
reads
$[ \fsim^{\mathrm{min}} - 1.96 \,\sigma_{\fexp}, \;
\fsim^{\mathrm{min}} + 1.96 \,\sigma_{\fexp} ]$,
where $\sigma_{\fexp}$ is a proxy for the standard deviation of 
$\fsim^{\mathrm{min}}$.
Substituting the upper bound of the confidence interval for 
$\fsim^{\mathrm{min}}$ into Eq.~(\ref{eq:def:mse}) 
yields an upper bound of the mean square error 
\begin{eqnarray}
E^{2}_+ &=& \frac{1}{N}\, \sum_{\ri,t}  
\frac{ \left(\fsim^{\mathrm{min}}(\ri,t) + 1.96*\sigma_{\fexp}(\ri,t)     -\fexp(\ri,t)\right)^{2}}{\sigma_{\fexp}^2(\ri,t)}
\label{eq:def:mse:bound:1} \\ 
&=& E^2_{\mathrm{min}} + 2 * 1.96 * 
\frac{1}{N}\, \sum_{\ri,t}  
\frac{ \fsim^{\mathrm{min}}(\ri,t)  -\fexp(\ri,t) }{\sigma_{\fexp}(\ri,t)}
+ 1.96^2
\label{eq:def:mse:bound:2}  
\end{eqnarray}
We define the confidence region for the parameters $(\gre,D)$ by
the domain within the level contour $E^{2}(\gre,D) = E^{2}_+ = 15.8$, see 
Fig.~3C. Conservative estimates of confidence intervals
on $\gre$ and $D$ (in brackets) are finally obtained by inscribing this contour
within a rectangle:
\begin{eqnarray}
\gre & = & 10\;\mu\mathrm{m^{2} h^{-1}} \quad[6,13]
\label{eq:CL:gamma1}\\
D & = & 1.6\;\mu\mathrm{m^{2} h^{-1}} \quad[0.5,3.9]
\label{eq:CL:D}
\end{eqnarray}

The value of the dimensionless ratio
$\frac{D_{2}}{\ri D_{1}}$  
allows  to determine
whether drift or diffusion dominate the dynamics. Since
\begin{equation}
  \label{eq:ratio:v2}
\left|  \frac{D_{2}}{\ri D_{1}} \right| =\frac{D}{\gre} = 0.16 \quad 
[0.04, 0.65]
\end{equation}
we conclude that drift dominates, but that diffusion cannot be neglected.

\subsection{Influence of an epithelial tension}
\label{sec:fit:sigma1}

Since tissues may quite generally be under compression or under tension,
we first tested the robustness of our results by taking into account
an epithelial tension $\snre$ of unknown sign.
Force balance is modified: a reduced tension coefficient
$\sre=\snre/\friction$  contributes to the drift coefficient.
Following the procedure given in Sec.\ref{sec:fit:leastsquares}, here 
based on the Langevin equation 
\begin{equation}
\dot{\rt}(t)=-\frac{\gre}{\rt}+ \sre + \sqrt{ 2D}
\,\eta(t)
\label{eq:Langevin:v4}
\end{equation}
the optima and confidence intervals 
for the three unknown parameters $(\gre,\sre,D)$  are
\begin{eqnarray}
\gre & = & 7\,\mu\mathrm{m^{2} h^{-1}} \, [0, 26]
\label{eq:withsigma:gamma1}\\
D & = & 1.5\,\mu\mathrm{m^{2} h^{-1}} \, [0.4, 3.9]
\label{eq:withsigma:D}\\
\sre & = & -0.15 \,\mu\mathrm{m \,h^{-1}} \, [-0.65, 0.65]
\label{eq:withsigma:sigma1}
\end{eqnarray}
for a minimal value of the error
$E^2_{\mathrm{min}} = \mathrm{min}_{\gre,\sre,D}\, E^2(\gre,\sre,D) = 7.2$
(see Supp. Fig.~7A).  

Strikingly, zero belongs to the confidence interval for $\sre$, and 
the level of agreement between predictions and data is not improved when
taking into account the (small, negative) optimal value 
(\ref{eq:withsigma:sigma1})
(compare Supp. Fig.~7AB with Fig.~3AB).
Further, the revised estimates (\ref{eq:withsigma:gamma1}-\ref{eq:withsigma:D})
for $\gre$ and $D$ are consistent with
the previous confidence intervals (\ref{eq:CL:gamma1}-\ref{eq:CL:D}).
We conclude that the closure fraction 
data is consistent with the absence of measurable tension 
in the monolayer ($\snre = 0$). Indeed little to no retraction was observed
when performing laser ablation in the monolayer away from the margin.

\subsection{Influence of fluctuations of the cable tension}
\label{sec:fit:gamma2}

The diffusion coefficient $D$ is proportional to the amplitude
of fluctuations of the epithelial tension about its (zero) average value.
We also tested the robustness of our results by taking into account 
possible fluctuations of the the cable tension, of amplitude 
$\dg \ge 0$. Assuming for simplicity that fluctuations
in the cable tension and in the epithelial tension are uncorrelated,
we express the diffusion $D_{2}(\rt,t)$ as 
\begin{equation}
D_{2}(\rt,t) = \ds + \frac{\dg}{\rt^{2}}
\label{eq:def:D2:Dgamma}
\end{equation}
and study the modified Langevin equation
\begin{equation}
\dot{\rt}(t)=-\frac{\gre}{\rt}+ 
\sqrt{ 2 \left( D + \frac{\dg}{\rt^{2}} \right)}
\,\eta(t)
\label{eq:Langevin:v5}
\end{equation}
interpreted according to Ito's rule.
This model leads to the estimates
\begin{eqnarray}
\gre & = & 10 \,\mu\mathrm{m^{2}.h^{-1}} \, [4, 12]
\label{eq:withgamma2:gamma1}\\
\ds & = & 1.5 \,\mu\mathrm{m^{2}.h^{-1}} \, [0, 2.5]
\label{eq:withgamma2:D}\\
\dg & = & 10 \,\mu\mathrm{m^{4}.h^{-1}} \, [0, 870]
\label{eq:withgamma2:gamma2}
\end{eqnarray}
for a minimal value of the error
$E^2_{\mathrm{min}} = \mathrm{min}_{\gre,\ds,\dg}\, E^2(\gre,\ds,\dg) = 7.5$.
We emphasize that: \emph{(i)} the optimal values
in (\ref{eq:withgamma2:gamma1}-\ref{eq:withgamma2:D}) are consistent
with the confidence intervals 
(\ref{eq:CL:gamma1}-\ref{eq:CL:D}); \emph{(ii)}  
a non-zero value of $\dg$ (\ref{eq:withgamma2:D}) has little influence on
the level of agreement between theoretical and experimental closure fractions
(compare Supp. Figs.~6C-D and Figs.~3A-B).

Taking into account fluctuations of the cable tension allows to 
define two critical radii:
\begin{equation}
  \label{eq:def:Rc}
  R_{\gamma}^{(1)} = \sqrt{\frac{\dg}{\ds}}, \quad \quad 
  R_{\gamma}^{(2)} = \sqrt{\frac{\dg}{\gre}}
\end{equation}
Fluctuations of the cable tension dominate  fluctuations of the 
epithelial tension below $R_{\gamma}^{(1)}$, and dominate the deterministic
cable tension below $R_{\gamma}^{(2)}$.
Using (\ref{eq:withgamma2:gamma1}-\ref{eq:withgamma2:gamma2}),
we find $R_{\gamma}^{(1)} \simeq 3 \, \mu$m and 
$R_{\gamma}^{(2)} \simeq 1 \, \mu$m:  this suggests that cable tension
fluctuations may dominate the very late stage of the closure process.

Conversely, we find that cable tension 
fluctuations are negligible except near closure, and
conclude by selecting the most parsimonious model, 
Eq.~(\ref{eq:Langevin:v2}), which we will use in Sec.~\ref{sec:quant}
to define and compute additional quantifiers of closure dynamics.

\section{Statistical quantifiers of closure dynamics}
\label{sec:quant}

\subsection{Closure half-time $\thalf$}
\label{sec:quant:half}

We first consider the closure half-time $\thalf$, defined as
the time needed to close half of the wounds for a given initial radius $\ri$: 
\begin{equation}
f(\ri,\thalf)=\frac{1}{2}.\label{eq:def:t12}
\end{equation}
Experimentally, $\thalf$ becomes larger than the total duration
of the experiment $\thalf>83$ h above
a radius $\ri= 44.5\,\mu\mathrm{m}$.
In Fig.~3D, we compare our measurement of 
the half-closure time up to $\ri= 49.5\,\mu\mathrm{m}$
with the outcome of numerical simulations for the optimal parameters
(\ref{eq:best:gamma1}-\ref{eq:best:D}).
Experimental error bars are obtained from  the maximum
$\mathrm{max\left(t^{+} - \thalf, \thalf - t^{-}\right)}$
where the times $t^{\pm}$ are defined by
$\fexp(\mathrm{\ri,t}^{\pm})=
\frac{1}{2}\left(1\pm\frac{1}{\nexp(\ri)}\right)$.
As expected by comparing Figs.~3A-B, where the same closure fraction
data and theoretical frequencies are plotted as a heat map, experimental
and theoretical values agree very well in Fig.~3D.

\subsection{Distribution of closure times $\tc$}
\label{sec:quant:disttc}

Figs.~2C-E compare experimental and theoretical
closure time distributions 
for radii $\ri=34$, $40$ and $46\,\mu\mathrm{m}$.
Experimental values are computed, within $\Delta \ri = 1.5\,\mu\mathrm{m}$ 
and binned over time intervals of $\Delta t = 12$h, 
by discrete differentiation of $\fexp$,
\begin{equation}
\label{eq:pdf}
 \pexp(\tc \mid \ri) =
\frac{\fexp(\ri,\tc+\Delta t)-\fexp(\ri,\tc-\Delta t)}{2\Delta t}
\end{equation}
The error bars in Figs.~2C-E 
are calculated from (\ref{eq:def:var:exp}) according to
\begin{equation}
\label{eq:dpdf}
\Delta \pexp(\tc \mid \ri)= 
\left(\frac{\sigma_{\fexp}^2(\ri,\tc+\Delta t) +
\sigma_{\fexp}^2(\ri,\tc-\Delta t)}{4 \Delta t^2}
\right)^{\nicefrac{1}{2}}.
\end{equation}
Theoretical frequencies are computed using Eq.~(\ref{eq:def:pdf:tc}).
We find reasonable agreement within error bars.

\subsection{Mean closure time as a mean first-exit time}
\label{sec:quant:meantc}

The mean closure time $\langle \tc(\ri) \rangle$
is defined as the mean first-exit time to zero starting 
from an initial radius $\rt(0) = \ri$ [30].
We show that it admits a simple analytical expression
for the stochastic process defined by the Langevin equation 
(\ref{eq:Langevin:v2})  with an absorbing boundary
at $r = 0$ and a reflecting boundary at $r = \ri$.

The solution of the differential equation
\begin{equation}
\label{eq:evol:T1}
-\frac{\gre}{x} \,\DT{}{x} T_1(x) + D \, \DTn{2}{}{x} T_1(x) = -1
\end{equation}
supplemented with the boundary conditions 
$T_1(x = 0)=0$, $\DT{}{x} T_1(x=\ri) =0$ reads
\begin{equation}
\label{eq:sol:T1}
T_1(x) = \frac{1}{\gre-D} \left(
\frac{x^2}{2} - 
\frac{x^{1 + \gre/D} \, \ri^{1 - \gre/D}}{1+\gre/D} 
\right)
\end{equation}
This yields the mean closure time as a function of initial radius
$\langle \tc(\ri) \rangle = T_1(x = \ri)$:
\begin{equation}
\label{eq:firstpassage:v2}
\langle \tc(\ri) \rangle=\frac{\ri^2}{2(\gre+D)} 
\end{equation}
This prediction is compared to experimental data in Fig.~2F.

In the absence of force fluctuations, Eq.~(\ref{eq:Langevin:v2}) reduces
to $\dot{\rt}(t)=-\frac{\gre}{\rt}$, with the solution
$\rt(t)^2 = \ri^2 - 2 \gre t$. For a given initial radius, the deterministic
closure time reads
 \begin{equation}
\label{eq:closure:det}
\tc^{\mathrm{det}}(\ri)=\frac{\ri^2}{2\gre}
\end{equation}
As a consequence the ratio
 \begin{equation}
\label{eq:closure:ratio}
\frac{\tc^{\mathrm{det}}(\ri)}{\langle \tc(\ri) \rangle}=1+\frac{D}{\gre}
\end{equation}
is always larger than unity: in the presence of fluctuations,
the mean closure time $\langle \tc(\ri) \rangle$ 
is always shorter than the deterministic closure time 
$\tc^{\mathrm{det}}(\ri)$.

\subsection{Fluctuations of the closure time}
\label{sec:quant:vartc}

Higher moments can be calculated iteratively [30]. 
In the case of the second moment $\langle \tc^2(r) \rangle$, 
we solve the differential equation:
\begin{equation}
\label{eq:var}
-\frac{\gre}{x} \,\DT{}{x} T_2(x)
+D \, \DTn{2}{}{x} T_2(x) = -2 T_1(x)
\end{equation}
with the boundary conditions $T_2(x = 0)=0$ 
and $\DT{}{x} T_2(x=\ri) =0$, and expression~(\ref{eq:sol:T1}).
The second moment $\langle \tc^2(\ri) \rangle = T_2(x = \ri)$ reads
\begin{equation}
\label{eq:firstpassage:m2}
\langle \tc^2(\ri) \rangle = 
\frac{5 + \gre/D}{(3 + \gre/D) \, (1+\gre/D)^2} \, 
\frac{\ri^4}{4 D^2} 
\end{equation}
The variance simplifies to
\begin{equation}
\label{eq:firstpassage:var}
 \sigma^2_{\tc}(\ri) = \langle \tc^2(\ri)\rangle - \langle \tc(\ri) \rangle^2
= \frac{2D}{3D+\gre} \, \langle \tc(\ri) \rangle^2
 \end{equation}
and the coefficient of variation of the closure time, 
defined as the ratio of the standard deviation to the mean value, 
is a constant
\begin{equation}
\label{eq:var:ratio}
 \frac{\sigma_{\tc}({\ri})}{\langle \tc(\ri) \rangle} =
 \left(\frac{2 D}{3D + \gre}\right)^{\frac{1}{2}} 
 \end{equation}
This prediction is compared to experimental data in Fig.~2G.

\section{Initial mean square deviation}
\label{sec:initmsd}

In order to explain the diffusive behavior of the mean square deviation 
observed at short time (See Fig.~2I) we define 
\begin{equation}
  \label{eq:def:Y}
Y(t)= \frac{\rt(0)-\rt(t)}{\sqrt{2D}}  
\end{equation}
and obtain by substitution in (\ref{eq:Langevin:v2}), 
the Langevin equation for $Y(t)$: 
\begin{equation}
  \label{eq:Langevin:v3} 
\dot{Y}=  \tilde{D_{1}} +\eta(t)
\end{equation}
with a drift coefficient 
$\tilde{D}_{1}=\frac{\gre}{\rt \sqrt{2D}}$. 
The distribution $p_{o}(Y,t)=p(Y,t \mid 0, 0)$ obeys the 
(forward) Fokker-Planck equation
\begin{equation}
\label{eq:FP:Y}
\DP{}{t}p_{o}(Y,t)=-\DP{}{Y} \left( \tilde{D_{1}} \, p_{o}(Y,t) \right)
+ \frac{1}{2}\DPn{2}{}{Y} p_{o}(Y,t).
\end{equation}
Introducing the scaling variable 
\begin{equation}
  Z=\frac{Y}{\sqrt{t}}
\label{eq:def:Z}
\end{equation}
and assuming that $p_{o}(Y,t)=\frac{1}{\sqrt{t}}\,G(Z)$,
Eq.~(\ref{eq:FP:Y}) becomes
\begin{equation}
\DTn{2}{G}{Z}+Z\,\DT{G}{Z} +G
= 2\,\sqrt{t}\,\frac{\mathrm{d}}{\mathrm{d}Z}[\tilde{D}_{1}\,G] 
\end{equation}
The scaling Ansatz (\ref{eq:def:Z}) is thus valid in the limit 
\begin{equation}
  \label{eq:scaling:cond}
 2\sqrt{t}\,\tilde{D}_{1}\ll Z 
\end{equation}
where the differential equation obeyed by $G(Z)$ simplifies to
\begin{equation}
\DTn{2}{G}{Z}  + Z\, \DT{G}{Z} + Z = 0
\label{eq:def:FP2:scaling6}
\end{equation}
With the boundary condition $\lim_{Z\to\infty}G(Z)=0$, 
the normalized solution of Eq.~(\ref{eq:def:FP2:scaling6}) is 
$G(Z)= \sqrt{\frac{2}{\pi}} \, e^{-Z^2/2}$.
Given the second moment of $Z$
\begin{equation}
\label{eq:Z:var}
  \langle Z^2 \rangle = \intop_{0}^{+\infty}Z^{2}\, G(Z)\, \mathrm{d}Z = 1
\end{equation}
we obtain
\begin{equation}
  \label{eq:dr0:calc:1}
\langle (\rt(0)-\rt(t))^{2} \rangle = 
2D \, \langle  Y^2\rangle = 
2D \, \langle  Z^2\rangle \, t = 2 D \, t 
\end{equation}
as expected for simple diffusion.

Since the scaling variable is typically close to $Z \sim 1$
according to Eq.~(\ref{eq:Z:var}), condition (\ref{eq:scaling:cond}) amounts to 
$t \ll \frac{D}{2} (\frac{\rt}{\gre})^2$. 
During the early stage of closure, we expect that $\rt(t) \simeq \rt(0)$,
and for $30 \le \rt(0) \le 50 \, \mu$m,
using the numerical values (\ref{eq:best:gamma1}-\ref{eq:best:D}) of $\gre$ 
and $D$, the above inequality will be respected when $t \ll 7$ h.
Fig.~2I confirms that the approximate scaling solution 
deriving from $G(Z)$ is indeed relevant for $t < 2$h.

Note that the scaling (\ref{eq:dr0:calc:1}) does not depend on a 
specific functional form of the drift $D_1(r,t)$, and  remains valid 
for the stochastic process defined in Sec.~\ref{sec:fit:sigma1} 
with a non-zero epithelial tension $\snre$, although in a
range (\ref{eq:scaling:cond}) that depends on the drift coefficient.

When the diffusion coefficient $D_2(r,t)$ depends on $r$ as in 
Sec.~\ref{sec:fit:gamma2}, we checked that the same diffusive scaling 
also holds for short time $t$ and small deviations $r(0)-r(t)$: the 
$\rt$-dependence of $D_2$ can be transformed away by an appropriate 
definition of $Y$ that generalizes (\ref{eq:def:Y}). 
However, the factor $2D$ in Eq.~(\ref{eq:dr0:calc:1}) is then replaced 
by a coefficient that depends upon the experimental distribution of radii.






\end{document}